\documentclass[twocolumn,showpacs,preprintnumbers,amsmath,amssymb]{revtex4}

\usepackage{graphicx}
\usepackage{dcolumn}
\usepackage{bm}
\usepackage{longtable}

\begin{document}

\preprint{APS/123-QED}

\title{
Experimental evidence of a natural parity state in $^{26}$Mg
and its impact to the production of neutrons for the s-process.} 

\author{C. Ugalde\footnote{Electronic address: cugalde@unc.edu}, A. Champagne, S. Daigle, 
C. Iliadis, R. Longland, J. Newton, and E. Osenbaugh}
\affiliation{Department of Physics and Astronomy, University of North Carolina, 
Chapel Hill, North Carolina 27599, USA and Triangle Universities Nuclear Laboratory, 
Durham, North Carolina 27708, USA} 

\author{J.A. Clark, C. Deibel, A. Parikh\footnote{Current address: 
Departament de F\'isica i Enginyeria Nuclear, Universitat Polit\'ecnica de
Catalunya, Barcelona, Spain}, P. Parker, and C. Wrede}
\affiliation{Wright Nuclear Structure Laboratory, Yale University, New Haven, 
Connecticut 06520, USA} 

\begin{abstract} 
We have studied natural parity states in $^{26}$Mg
via the $^{22}$Ne($^{6}$Li,d)$^{26}$Mg reaction. Our method significantly
improves the energy resolution of previous experiments and, as a result,
we report the observation of a natural parity state in $^{26}$Mg.
Possible spin-parity assignments are suggested on the basis of published
$\gamma$-ray decay experiments. The stellar rate of the 
$^{22}$Ne($\alpha$,$\gamma$)$^{26}$Mg reaction is reduced and may  
give rise to an increase in the production of s-process neutrons via 
the $^{22}$Ne($\alpha$,n)$^{25}$Mg reaction.

\end{abstract}

\date{\today}
             
\maketitle
\section{Introduction}
The $^{22}$Ne($\alpha$,n)$^{25}$Mg reaction is regarded as the main 
neutron source for the s-process in core He-burning massive stars \cite{Hoffman:2001} and 
is of relevance in He-shell burning in AGB stars \cite{Straniero:1995}. Our current understanding 
of its rate is one of the most important sources of uncertainty in the
nucleosynthesis of heavy elements. 

\begin{figure}
\includegraphics[width=8.6cm]{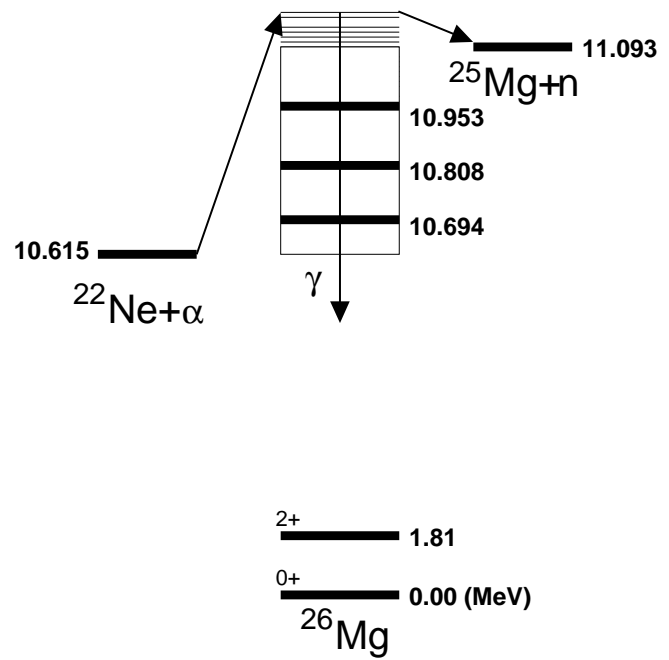}
\caption{\label{fig:levelsMg26}Level scheme (not drawn to scale)
of $^{26}$Mg showing the $^{22}$Ne + $\alpha$ entrance channel and the
two competing exit channels $^{25}$Mg + n and $^{26}$Mg + $\gamma$. 
The levels of interest to this work (all below the neutron
threshold) are shown. Note the negative Q-value 
(Q = -478.3$\pm$0.04 keV) for the $^{22}$Ne($\alpha$,n)$^{25}$Mg process.}
\end{figure}

The species $^{22}$Ne is produced in helium-rich environments
from $^{14}$N, a product of the CNO cycle, via 
$^{14}$N($\alpha$,$\gamma$)$^{18}$F($\beta^{+}$$\nu$)$^{18}$O($\alpha$,$\gamma$)$^{22}$Ne.
The production of neutrons by $\alpha$-particle capture on $^{22}$Ne occurs 
through a resonant process involving the formation of the $^{26}$Mg 
compound nucleus in an excitation range of high level density. The 
populated resonant levels decay by neutron emission to the
ground state of $^{25}$Mg. Figure \ref{fig:levelsMg26} shows both 
the formation channel (open above $E_x$=10,614.78$\pm$0.03 keV) and the outgoing 
neutron channel (open above $E_x$=11,093.07$\pm$0.04 keV). 

The $^{22}$Ne($\alpha$,$\gamma$)$^{26}$Mg reaction competes with the 
$^{22}$Ne($\alpha$,n)$^{25}$Mg process \cite{The:2007} above the neutron threshold.
At temperatures of relevance to the main and weak components of the 
s-process, the neutron yield is defined by both the abundance 
of $^{22}$Ne and the branching ratio between the competing $\gamma$-ray 
and neutron exit channels\cite{Woosley:2002}. Direct measurements of both processes 
have been reported by several groups for $E^{lab}_{\alpha}>$ 800 keV
(For example, see \cite{Jaeger:2001,Drotleff:1993,Giesen:1993,Harms:1991,Wolke:1989} 
and other references therein). However, due to the Coulomb 
barrier, the cross section remains uncertain for lower energies. 
A direct measurement for $E^{lab}_{\alpha}<$ 800 keV still needs to be done.

There is only scarce information on levels of $^{26}$Mg between the 
$\alpha$-particle and neutron thresholds. Giesen {\it et al.} \cite{Giesen:1993} 
report two natural parity resonances (at $E_x$=10.694(20) MeV 
and $E_x$=10.949(25) MeV) in 
their $^{22}$Ne($^{6}$Li,d)$^{26}$Mg experiments. As discussed by 
Karakas {\it et al.} \cite{Karakas:2006}, the main source of uncertainty in the rate for 
the $^{22}$Ne($\alpha$,$\gamma$)$^{26}$Mg reaction at these energies 
results from the unknown spin of the $^{26}$Mg state at $E_x$=10.949(25) MeV. 
However, in their calculation of the rate, Karakas {\it et al.} have considered 
only two states while more than 20 levels are listed in the compilation 
of Endt \cite{Endt:1998} for excitation energies between the $\alpha$-particle
and neutron emission thresholds. A detailed study of 
levels in $^{26}$Mg and their spins and parities at these energies 
is urgently needed. Here we report the first experimental step towards a 
complete understanding of the reaction rate at stellar temperatures.

Both $^{22}$Ne and $^{4}$He have a ground state with J$^{\pi}$=0$^{+}$.
Thus, preferentially natural parity states in  $^{26}$Mg can be populated via the 
$^{4}$He+$^{22}$Ne process \cite{Koehler:2002}. We studied the astrophysically relevant 
natural parity states in $^{26}$Mg between E$_x$=10615 keV and 
E$_x$=11093 keV via the $^{22}$Ne($^{6}$Li,d)$^{26}$Mg reaction, 
which preferentially populates natural parity states for direct 
$\alpha$-particle transfer \cite{Fulbright:1979}.

\section{Experiment}

Five $^{22}$Ne targets were prepared by implanting $^{22}$Ne into 40 
$\mu$g/cm$^2$, 99.9\% $^{12}$C-enriched foils. The foils were floated from glass 
slides in deionized water and mounted onto aluminum frames. It is thought 
that the mechanical stability of thin foils can be improved by exposing 
them to an intense burst of light \cite{Rowton:1995}. 
We flashed the unimplanted foils with 
a photographic strobe and observed their slackening. The $^{22}$Ne 
beam was produced by the 200 keV Eaton Ion Implanter at the 
University of North Carolina; two energies (20 keV and 35 keV) were used 
to implant both sides of the foils. $^{22}$Ne was implanted in this way to 
achieve a total density of 19-22$\times$10$^{16}$ atoms/cm$^{2}$ 
($\sim$ 7 $\mu$g/cm$^{2}$). The dose of implanted ions was estimated 
by integrating the beam current at the target, which together with the beam stop 
and the implantation chamber, acted as a Faraday cup. Secondary electrons were
suppressed with a negative voltage applied to a copper pipe placed in front of 
the target and coaxial to the beam. The copper pipe was cooled with liquid 
nitrogen to prevent natural carbon build up on the targets. It was important to keep 
$^{13}$C contamination of the targets to a minimum because deuterons from the 
$^{13}$C($^{6}$Li,d)$^{17}$O reaction would have posed a major source of background 
in our $^{22}$Ne($^{6}$Li,d)$^{26}$Mg spectra. An additional source of background 
could have been $^{20}$Ne from the ion source at the implanter. The natural 
neon gas used to produce the beam was mass analyzed using a magnet with a mass 
resolution better than $\Delta$M/M=0.01 allowed us to get an excellent separation 
of $^{22}$Ne and $^{20}$Ne. The $^{22}$Ne beam current was kept below 400 nA to 
minimize heating of the carbon foils. 

Both the $^{22}$Ne($^{6}$Li,d)$^{26}$Mg experiment and analysis of the target 
composition were performed with a $^{6}$Li beam produced by the Wright 
Nuclear Structure Laboratory ESTU Tandem van de Graaff accelerator at Yale 
University. The reaction products were momentum-separated with the Enge 
split-pole spectrometer and detected at the focal plane with a position-sensitive 
gas ionization chamber and a scintillator \cite{Parikh:PhDThesis}. 

The detector setup allowed us to separate different particle groups 
by means of the magnetic rigidity ($B \rho$), the energy 
loss ($\Delta E$) measured from the cathode and 
the residual energy ($E$) deposited at the scintillator. The magnetic 
rigidity ($B \rho$) was derived from independent position signals 
along two parallel wires (front wire and rear wire). The horizontal
acceptance was $\pm$10.7 mrad, which corresponded to the minimum 
aperture available.

\begin{figure*}
\includegraphics[width=17.5cm]{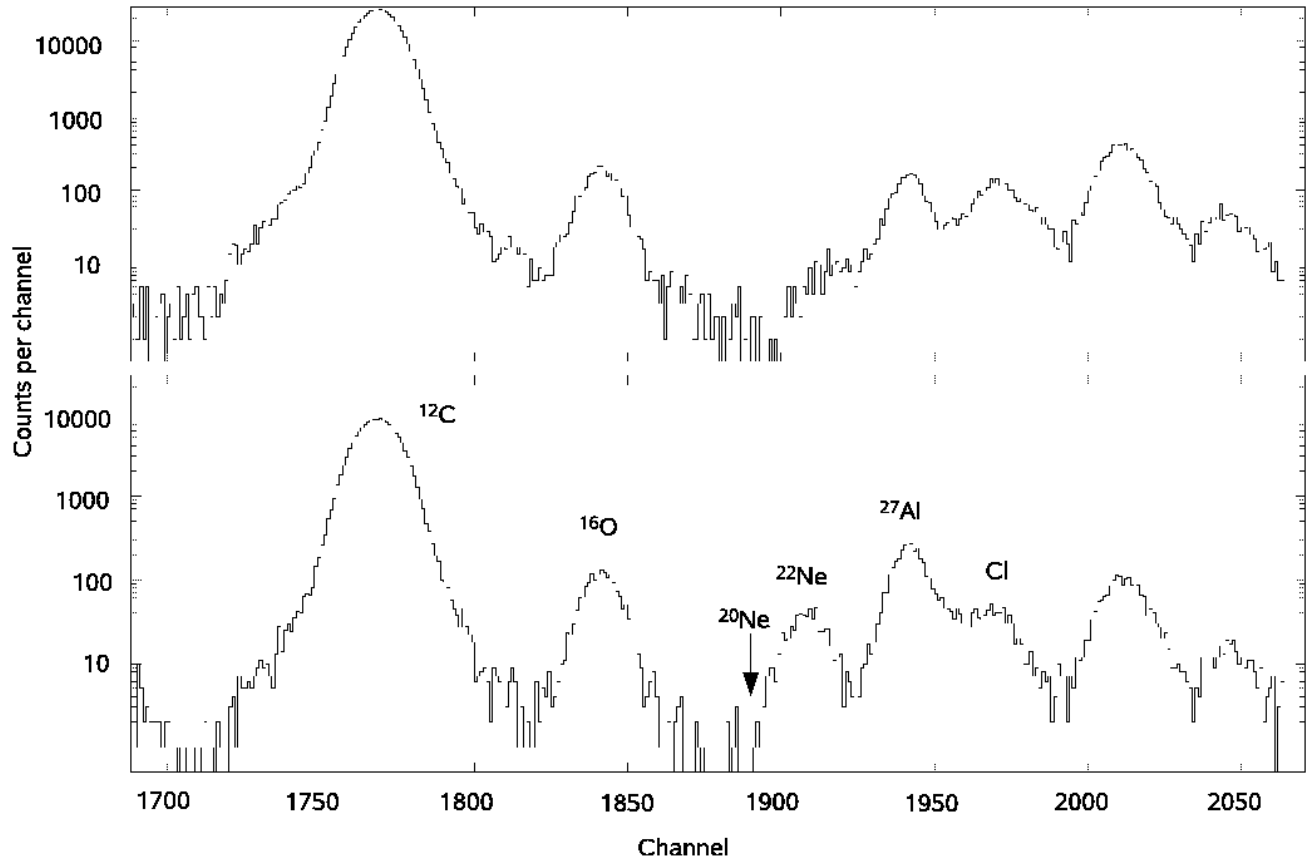}
\caption{\label{fig:elastic}
Elastic scattering off a non-implanted $^{12}$C substrate (upper panel) and
off a $^{22}$Ne-implanted $^{12}$C substrate (lower panel). The arrow shows 
the position where contaminating $^{20}$Ne would have appeared.
}
\end{figure*}

The target content analysis was performed with a 7.7 MeV $^6$Li beam.
To calibrate the elastic scattering position spectrum as a function 
of mass, we used a target consisting of a SiO$_2$ layer deposited 
on a 40$\mu$g/cm$^2$ natural carbon substrate. Elastic scattered 
$^6$Li off the target was observed in three major groups, each 
corresponding to $^{12}$C, $^{16}$O, and $^{28}$Si. A $^{22}$Ne-implanted 
and a non-implanted $^{12}$C foil 
were exposed to a $^6$Li beam as well (see figure \ref{fig:elastic}). 

The ($^6$Li,d) experiment was performed with a 30 MeV $^6$Li beam.
We took spectra with the non-implanted target for magnetic 
rigidity calibration purposes and for comparison with the 
$^{22}$Ne-implanted target. Because the cross section is 
expected to be larger at small scattering 
angles, we placed the spectrometer at 6$^o$ in the laboratory.
The spectrometer field was set to center the deuterons from the 
$^{12}$C($^{6}$Li,d)$^{16}$O reaction at the position spectrum.

\begin{figure*}
\includegraphics[width=17.5cm,height=13.5cm]{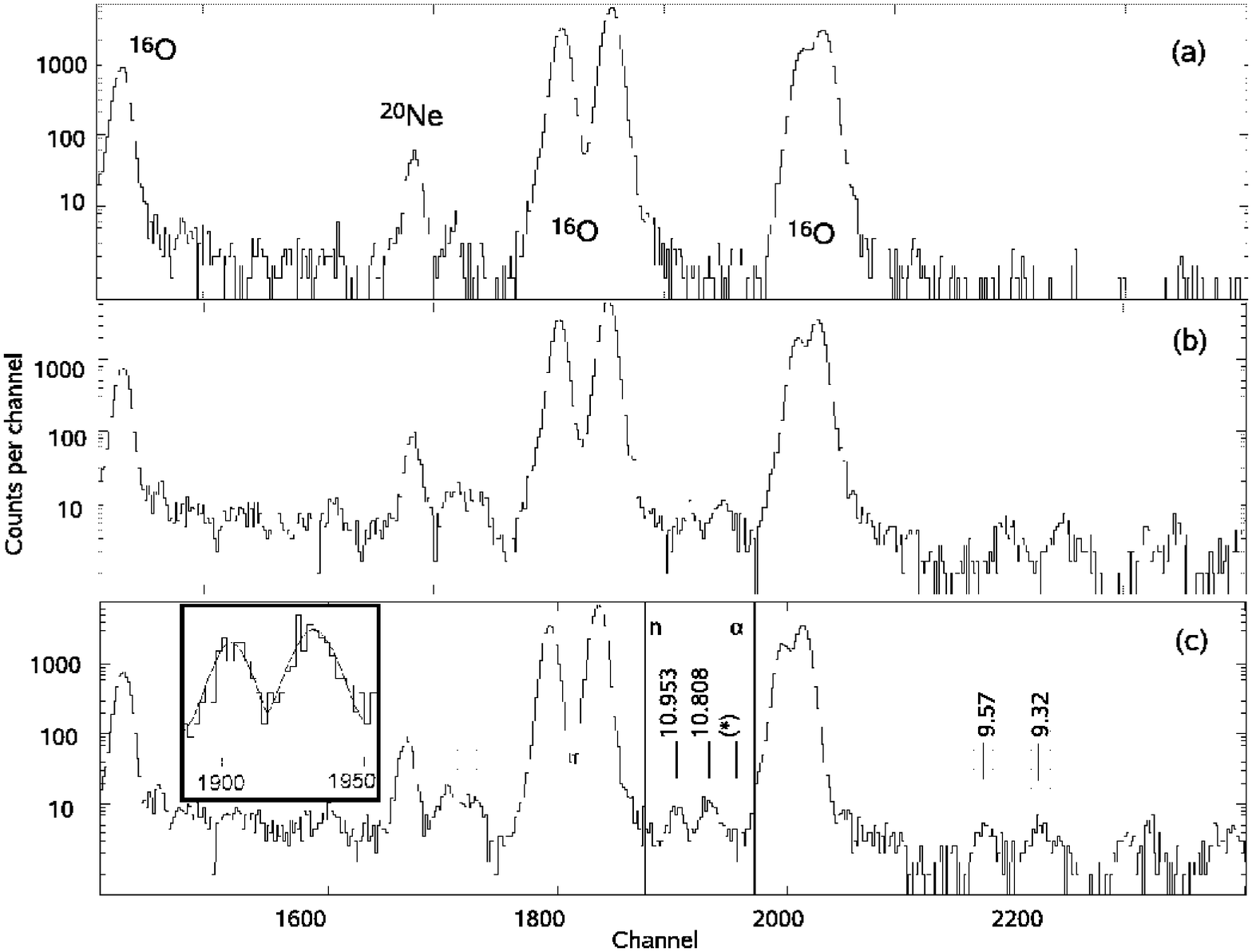}
\caption{\label{fig:deuteronspectra}
Deuteron position spectra for $\theta_{lab}$=6$^o$. (a) The top panel shows the deuterons from the $^{12}C$ substrate. 
Six groups are observed and correspond, from left to right, to the $E_x=8.8719(5)$ MeV state in 
$^{16}$O (from $^{12}$C($^{6}$Li,d)$^{16}$O), the $E_x=5.6214(17)$ MeV state in 
$^{20}$Ne (from $^{16}$O($^{6}$Li,d)$^{20}$Ne), the  $E_x$=7.11685(14) and 6.9171(6) MeV doublet in 
$^{16}$O, and the $E_x$=6.12989(4) and 6.0494(10) MeV doublet in $^{16}$O, both from 
$^{12}$C($^{6}$Li,d)$^{16}$O. (b) The middle panel shows the deuteron front position spectrum 
from a $^{22}$Ne-implanted $^{12}$C substrate. The energy region of interest in this 
work is the window located between the two $^{16}$O doublets. Deuteron groups not observed 
with the $^{12}$C target are seen here. (c) The lower panel shows the deuteron spectrum 
after reconstructing the focal plane for the $^{22}$Ne-implanted $^{12}$C substrate 
(see section \ref{section:analysis}). 
The energies of some of the observed $^{26}$Mg states are shown. The two vertical lines 
correspond, from left to right, to the positions of the neutron and $\alpha$-particle thresholds. 
The arrow labeled with an asterisk (*) corresponds to $E_x$=10.694 MeV. A state was not observed 
at this energy and angle (see text for discussion). The inset shows the Gaussian fits to the two 
peaks in the region between the neutron and $\alpha$-particle thresholds.
}
\end{figure*}

We acquired deuteron position spectra with an average 
beam current of 80 pnA with the $^{22}$Ne target and a total 
solid angle of 1.5 msr. The result is 
shown in figure \ref{fig:deuteronspectra}. 
The energy region of interest for deuteron groups corresponding to $^{26}$Mg 
states is located between the two $^{16}$O doublets. Deuteron groups unobserved with the 
$^{12}$C target appear both inside and outside the region of interest.

\section{Analysis and results}
\label{section:analysis}
\begin{figure}
\includegraphics[width=8.6cm,height=7cm]{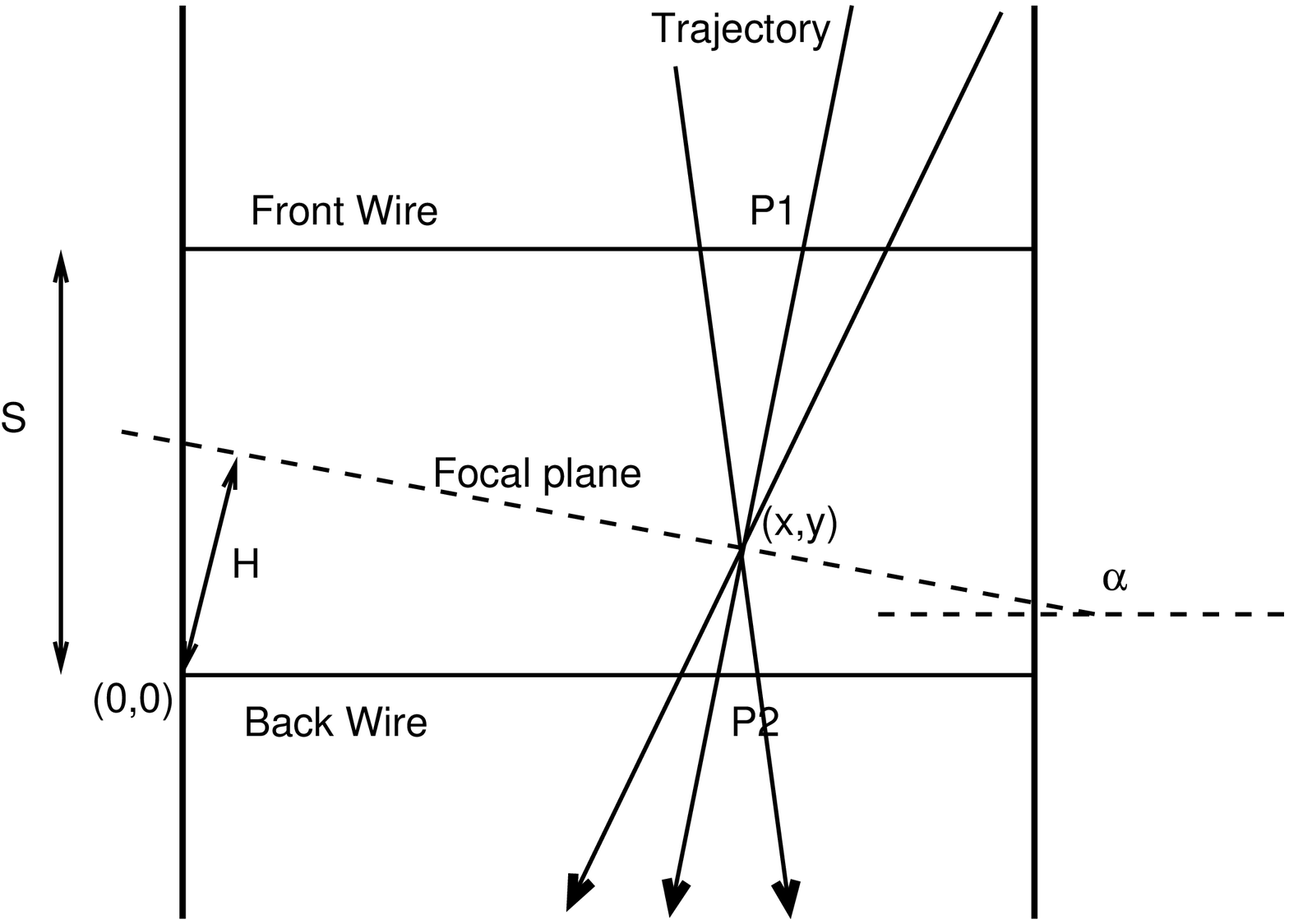}
\caption{\label{fig:detector}
Deuteron trajectory and focal plane diagram (adapted from \cite{Shapira:1975}).
}
\end{figure}

Although the detector was positioned so that the focal plane coincided nominally 
with the front wire, the two independent position measurements allowed us to 
more accurately locate the true focal plane by reconstructing the particle 
trajectories \cite{Shapira:1975}, thereby improving the position resolution.
Let $P_1$ and $P_2$ be the positions measured at the front and rear wire, 
respectively, and S the distance between the wires ($S=2.5$ cm). The trajectory 
$(x,y)$ of particles traveling in a plane parallel to the two wires is described 
by the relation
\begin{equation}
(P_1-P_2)y+Sx-P_2S=0.
\end{equation}
On the other hand, the $(x,y)$ equation for the focal plane is
\begin{equation}
{{x}\over{(1+\cot^2 \alpha)^{{1 \over 2}}}}+{{y}\over{(1+\tan^2 \alpha)^{{1 \over 2}}}}-H=0,
\end{equation}
where $\alpha$ is the angle between the focal plane and the front wire, and H is the
distance from the focal plane to the origin $(0,0)$. Solving $(x,y)$ simultaneously 
for the two equations one gets the position of the particles at the focal plane 
(see figure \ref{fig:detector}). The best position resolution 
was obtained for S/H=2. The resulting spectrum is shown in 
figure \ref{fig:deuteronspectra}(c); here the energy resolution improved from 
80 keV to 58 keV , as measured for the $^{16}$O peak at E$_x$=7.12 MeV. 

The S/H value was chosen to optimize the resolution of the $^{16}$O peaks in the 
position spectrum. The main source of background was the $^{12}$C substrate in 
the target, as can be noted by comparing panels (a) and (c) in 
figure \ref{fig:deuteronspectra}. Therefore, the best peak to 
background ratio for the $^{26}$Mg states was 
obtained by focusing on the $^{16}$O peaks instead of the $^{26}$Mg states 
themselves. Nevertheless, the width of $^{26}$Mg peaks for S/H=2 was still 
within 5\% of the optimized value for the focused $^{26}$Mg peaks (S/H=1.67).  
The width observed for the $^{26}$Mg peaks is listed in table \ref{tbl:calibration}.

\begin{figure}
\includegraphics[width=8.6cm,height=7cm]{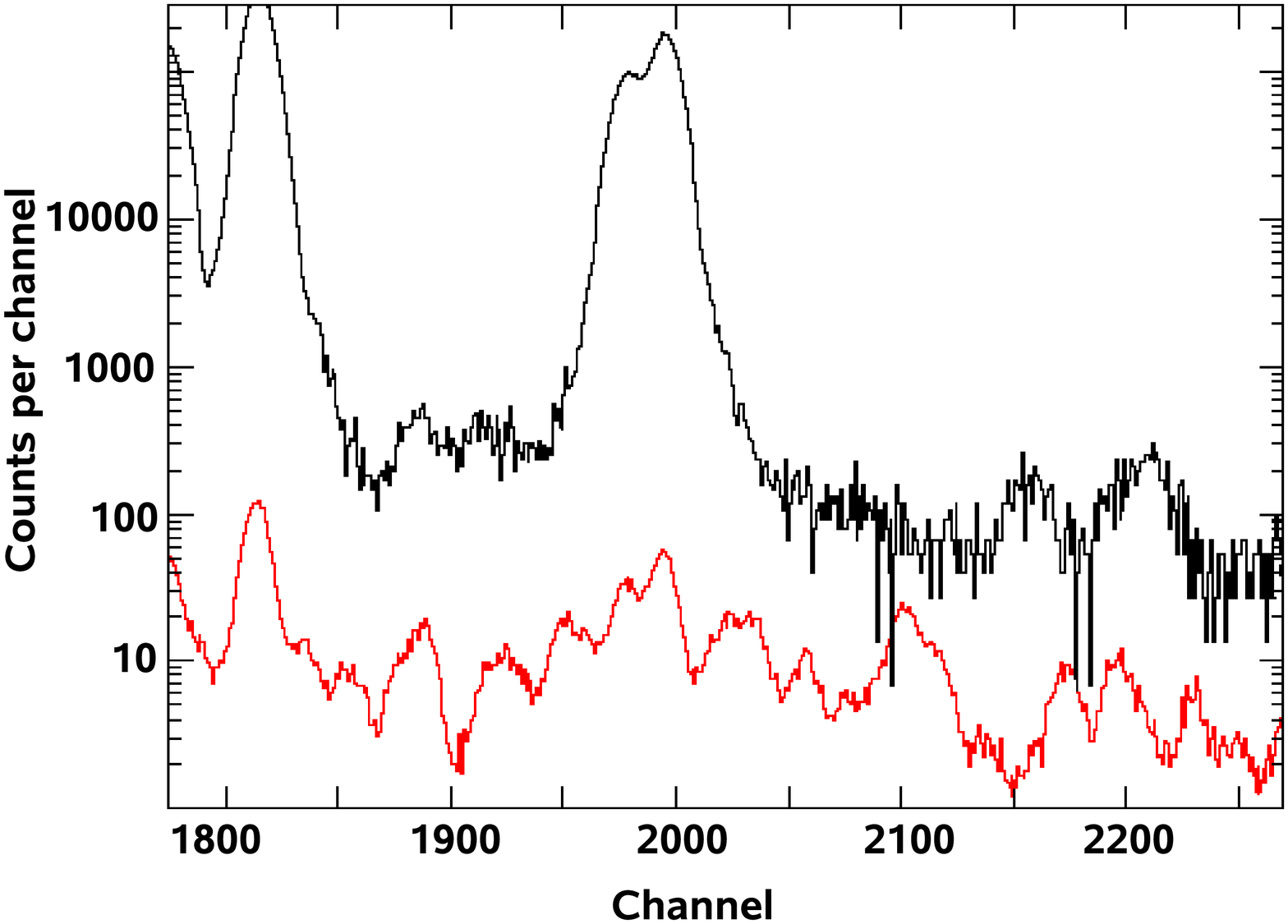}
\caption{\label{fig:alspectrum}
(Color figure). Comparison between spectra obtained with both a $^{22}$Ne-implanted target (black curve)
and an $^{27}$Al target (red curve). The yield for the $^{27}$Al target was renormalized to match the 
relative accumulated charge on the $^{22}$Ne-implanted target and the relative $^{27}$Al content in the targets, 
as measured with elastic scattering of $^6$Li. Both spectra were taken with a beam energy 
of 30 MeV and the spectrometer placed at 6$^o$. 
}
\end{figure}

The origin of the deuteron groups observed with the $^{22}$Ne target was established 
on the basis of a target content analysis via $^{6}$Li elastic scattering. 
The target content analysis was performed by moving the spectrometer to 20$^o$ in the
laboratory frame and reducing the magnetic field to 7.7 kG, so that elastically scattered
$^{6}$Li ions were centered at the front position spectrum.  The results for both the 
non-implanted and $^{22}$Ne-implanted $^{12}$C substrates are shown in 
figure \ref{fig:elastic}. The major peak appearing in the spectra 
corresponds to the ground state in 
$^{12}$C. Other elastic scattering groups were also identified. A comparison 
between these two elastic scattering spectra shows that 
the only group observed in the $^{22}$Ne-implanted target and not observed in 
the non-implanted target is that corresponding to $^{22}$Ne. The position where 
$^{20}$Ne contamination from the implantation process would have appeared is 
marked as well. From figure \ref{fig:elastic} it is clear that $^{27}$Al 
is also increased relative to the $^{12}$C group between both targets (probably 
as a result of the implantation process).  Therefore, we compared 
the deuteron spectrum obtained from an $^{27}$Al target with 
that of the $^{22}$Ne-implanted target. We scaled the background 
with the $^{27}$Al content determined via elastic scattering 
in the $^{22}$Ne-implanted target (see figure \ref{fig:alspectrum} 
for a content-normalized comparison between spectra 
from both targets). We concluded the peaks 
observed between the two $^{16}$O doublets originated from the
$^{22}$Ne($^{6}$Li,d)$^{26}$Mg reaction, with a contribution of 5$\%$
from $^{27}$Al to the area under the peaks. 

From the position spectrum, a magnetic rigidity and an excitation
energy calibration can be obtained by assuming a polynomial relation for the 
radius of curvature $\rho$. This requires that at least three peaks be unambiguosly 
identified. Here we used six states that were populated via the 
$^{12}$C($^{6}$Li,d)$^{16}$O and $^{16}$O($^{6}$Li,d)$^{20}$Ne 
reactions (see table \ref{tbl:calibration}).

The position of peaks observed was determined by fitting with a Gaussian template. The error bars 
of centroid positions were determined by sampling the area and width of the Gaussian determined 
by minimizing the value of $\chi^2$ for the fit, within error bars. 
A Monte Carlo sampling produced a set of centroid 
positions that determined the size of error bars. Table \ref{tbl:calibration} shows the 
centroids of the fitted peaks and the published excitation energies \cite{Firestone:1995} 
for the six calibration peaks. We fitted $\rho$, the deuteron trajectory's 
radius of curvature, as a function of the front wire position $P_1$ with an 
expression of the type
\begin{equation}
\rho(P_1)=a_0+ a_1(P_1-P_1[0])+...
\end{equation}

\begin{table}
\caption{\label{tbl:calibration}States observed in this work}
\begin{ruledtabular}
\begin{tabular}{ccccccc}
Centroid & $E_x$ & E$_{d}$\footnotemark[3]   & Nucleus & Peak width\\
(channel) & (MeV) & (MeV) &   & (keV) \\
\hline
1419.6(10)   &  8.8719(5)\footnotemark[1]	& 24.147(2)   & $^{16}$O  & \\
1667.1(10)   &  5.7877(26)\footnotemark[1]	& 25.450(11)  & $^{20}$Ne & \\
1792.4(15)   &  7.11685(14)\footnotemark[1]	& 26.122(1)   & $^{16}$O  & \\
1834.4(10)   &  6.9171(6)\footnotemark[1]       & 26.345(2)   & $^{16}$O  & \\

1903.2(50)    &  10.953(25) \footnotemark[2] 	& 26.719(61)  & $^{26}$Mg & 58(16)\\
1931.8(40)    &  10.808(20) \footnotemark[2]	& 26.872(50)  & $^{26}$Mg & 69(16)\\

1996.8(20)    &  6.12989(4)\footnotemark[1]     & 27.224(1)   & $^{16}$O  & \\
2012.6(20)    &  6.0494(10)\footnotemark[1]     & 27.314(5)   & $^{16}$O  & \\

2173.5(80)    &  9.57(4) \footnotemark[2]	& 28.19(12)   & $^{26}$Mg & 117(15)\\
2221.4(110)   &  9.32(6)\footnotemark[2]	& 28.46(18)   & $^{26}$Mg & 172(20)\\

\end{tabular}
\footnotetext[1]{From \cite{Firestone:1995}, used as calibration peaks.}
\footnotetext[2]{This work.}
\footnotetext[3]{Deuteron energy, as calculated from a kinematic analysis.}
\end{ruledtabular}
\end{table}

Here the $a_i$'s are the parameters of the fit.
We found that the best fit ($\chi^2$/N=0.1, where N=3 is the number of degrees of freedom) was 
obtained with a polynomial of degree 2. Finally, the excitation energy of $^{26}$Mg 
states was computed from $\rho$ with a kinematic analysis. The error bars  
include contributions from the uncertainty in the 
position of the centroid of the peaks and from the energy calibration.

\section{Discussion}
The recent calculation of the rate for 
the $^{22}$Ne($\alpha$,$\gamma$)$^{26}$Mg reaction \cite{Karakas:2006} 
includes contributions from two $^{26}$Mg states below the neutron threshold.
The largest source of uncertainty is from the 
$E_x$=10.949(25) MeV state; a second state at $E_x$=10.694(20) MeV 
(E$_{cm}$=0.078 MeV) has a negligible effect on the rate, as it 
is located too far below the Gamow window for He-burning temperatures 
(the window spans from  E$_{cm}$=0.36 to 0.57 MeV 
at T=2.0$\times$10$^8$ K). Both states were observed by Giesen et 
al. \cite{Giesen:1993} using the $^{22}$Ne($^{6}$Li,d)$^{26}$Mg 
reaction. However, we find evidence that their $E_x$=10.949(25) 
MeV state corresponds to at least two states in $^{26}$Mg.

Four $^{26}$Mg states were identified in this experiment (see 
table \ref{tbl:calibration}). Two states fall in the region between 
the two $^{16}O$ doublets (see figure \ref{fig:deuteronspectra}).
The first has $E_x$ = 10.808(20) MeV and the second has 
$E_x$ = 10.953(25) MeV. 

Giesen {\it et al.} \cite{Giesen:1993} and Giesen \cite{Giesen:Thesis} 
studied the $^{22}$Ne($^{6}$Li,d)$^{26}$Mg reaction 
with a relatively poor resolution of 120 keV in their deuteron 
spectra. In contrast, for the present experiment the energy 
resolution was 63 keV in the region of interest, as a result of using a solid target. 
We observed two $^{26}$Mg states outside of the region of interest. The first, at 
E$_x$=9.32(6) MeV, corresponds to the E$_x$=9.404(20) MeV state of Giesen {\it et al.}, 
while the second, observed at E$_x$=9.57(4) MeV, is in agreement to their state 
at E$_x$=9.586(20) MeV.

A state at $E_x$=10.694(20) MeV was not observed in our experiment, 
which is consistent with the $\theta_{lab}$=7.5$^o$ spectrum 
of Giesen \cite{Giesen:Thesis}. There he reports 
a cross section more than one order of magnitude smaller than that of the 
$E_x$=10.949(25) MeV state at the same angle; no peak 
corresponding to the $E_x$=10.694(20) MeV state in $^{26}$Mg can be identified
in the spectrum. Nevertheless, one single broad peak was observed at a higher 
excitation energy ($E_x$=10.949(25) MeV). With improved resolution we 
have resolved this peak into two $^{26}$Mg states. It is thus likely 
that the spectroscopic factor for this doublet gets contributions 
from both states, one of them (the state with $E_x$ = 10.808(20) MeV) is  
at an energy too low (E$_{cm}$=0.193 MeV) to contribute 
significantly to the reaction rate of $^{22}$Ne($\alpha$,$\gamma$)$^{26}$Mg
in the region of astrophysical interest.

We identify these two states as follows: Walkiewicz {\it et al.} \cite{Walkiewicz:1992} 
studied the secondary $\gamma$-rays 
from thermal neutron capture for $^{26}$Mg and observed an 8996.5 keV transition 
to the first excited state (2$^+$) of $^{26}$Mg. They report a state with 
$E_x$=10.8059(4) MeV and Endt \cite{Endt:1998} assigned J$^\pi$=(0$^+$-4$^+$) to it. 
This is consistent with our state at $E_x$=10.808(20) MeV. The 
$^{22}$Ne($^{6}$Li,d)$^{26}$Mg reaction most likely populates states 
with natural parity, so we propose this state to have 
$J^\pi$=0$^+$, 1$^-$, 2$^+$, 3$^-$ or 4$^+$.

Giesen {\it et al.} suggested the $E_x$=10.694(20) MeV state to have 
$J^\pi$=7$^-$, 8$^+$ or 4$^+$. Glatz {\it et al.} \cite{Glatz:1986} observed this same state to decay
by $\gamma$-ray emission to the $J^\pi$=5$^+$ state at E$_x$=7.395(1) MeV. Most likely, 
this decay corresponds to an M1 transition. This is in agreement with the $J^\pi$=4$^+$ 
assignment suggested by Karakas {\it et al.} \cite{Karakas:2006}.

An $E_x$=10.81(6) MeV $^{26}$Mg state was reported by 
Crawley {\it et al.} \cite{Crawley:1989} as well. They measured forward-angle cross 
sections for 201 MeV proton inelastic scattering and observed a
forward-peaked angular distribution, thus suggesting  $J^\pi$=1$^+$.
This state would not have been observed in the present experiment and
most likely corresponds to the $E_x$=10.824(3) MeV state observed 
by Moss \cite{Moss:1976} and listed in Endt \cite{Endt:1998}. 

\begin{figure*}
\includegraphics[width=17.5cm]{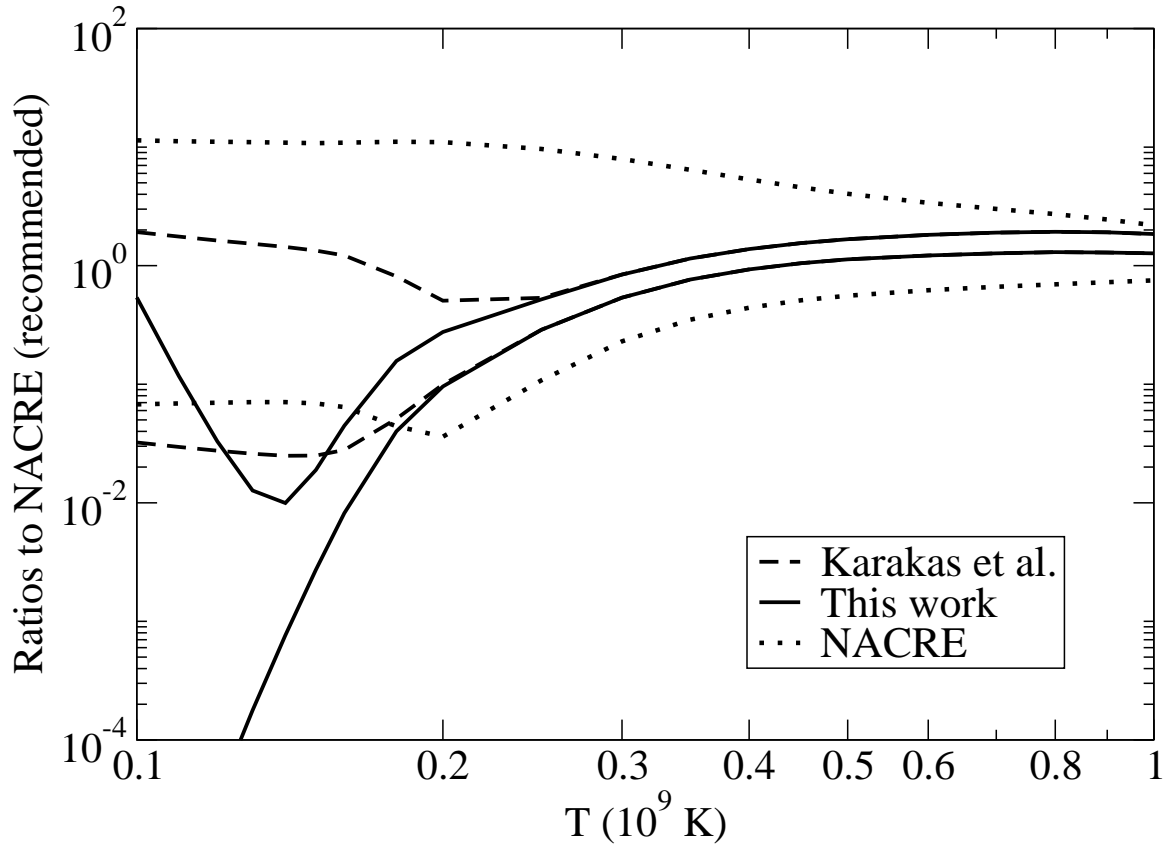}
\caption{\label{fig:ne22ag_ratios}
Ratio of upper and lower limits to the NACRE recommended 
rate \cite{Angulo:1999} for the  
$^{22}$Ne($\alpha$,$\gamma$)$^{26}$Mg reaction, as calculated by
Karakas {\it et al.} \cite{Karakas:2006} and this work. 
}
\end{figure*}

Our $E_x$=10.953(25) MeV state is at an energy consistent 
with the $E_x$=10.945(3) MeV level listed in Endt \cite{Endt:1998}.
The state at $E_x$=10.953(25) MeV was also observed by  
Giesen {\it et al.} \cite{Giesen:1993} and they suggested $J^\pi$=3$^-$ 
without being able to discard the $J^\pi$=2$^+$ and 4$^+$ 
assignments. Glatz {\it et al.} \cite{Glatz:1986} also observed this
state to decay by $\gamma$-ray emission to the $E_x$=8.625(1) MeV, 9.169(1) MeV,
and 9.383(1) MeV states with branching ratios 29(4), 61(5), and 10(2), respectively.
Endt \cite{Endt:1998} lists the three final states to have $J^\pi$=5$^-$, 6$^-$, 
and 6$^+$ respectively. Giesen {\it et al.}'s $J^\pi$ assignments are not consistent with 
these $\gamma$-ray decays. Assuming our experimental work
populated natural parity states in $^{26}$Mg, only $J^\pi$=5$^-$, 6$^+$, 
or 7$^-$ are allowed. The discrepancy between these results and Giesen {\it et al.}'s
probably comes from the fact that their DWBA analysis and J$^\pi$ assignment were 
performed for a peak consisting of two unresolved states.

We calculated the contributions from the two resolved resonances 
to the ($\alpha$,$\gamma$) rate by taking the total differential 
cross section observed by \cite{Giesen:1993} for their  $E_x$=10.949(25) MeV 
state and splitting it into two parts. The ratio of the two 
contributions was taken to be equal to the ratio of 
the areas under the peaks observed in our deuteron spectrum 
(Fig. \ref{fig:deuteronspectra}) after correcting for the background from 
$^{27}$Al, as shown in figure \ref{fig:alspectrum}. The 
spectroscopic factor S$_\alpha$  for each contribution was calculated 
by assuming a combination of spin pairs and then
by fitting a DWBA model computed with the code DWUCK4 \cite{Kunz:1983} to
the individual experimental cross sections. For the state at E$_x$=10.808(20) MeV,
S$_\alpha$=1.9$\times$10$^{-2}$, while at E$_x$=10.953(25) MeV, 
S$_\alpha$=2.8$\times$10$^{-3}$. The upper limit 
of the reaction rate was evaluated by assigning the states at
E$_x$=10.808(20) MeV and E$_x$=10.953(25) MeV as J$^\pi$=0$^+$ and 
J$^\pi$=5$^-$, respectively. For the lower limit we took 
J$^\pi$=4$^+$ and J$^\pi$=7$^-$, respectively.

Comparisons of our rates and the rates calculated by 
Karakas {\it et al.} \cite{Karakas:2006} to NACRE \cite{Angulo:1999} are shown in figure 
\ref{fig:ne22ag_ratios}. Relative to the NACRE rates, the
main effect is a reduction of the rate for temperatures 
below 0.3 GK. Thus an increase in the neutron production by the
($\alpha$,n) reaction may be expected in 
these stellar environments. Nevertheless, it is necessary to 
finalize the spin-parity assignment of all $^{26}$Mg states
contributing to the ($\alpha$,$\gamma$) rate and that are located 
below the neutron threshold. Further experimental studies are under way.

\section{Acknowledgements}
We wish to thank Adam Cottrel of the University of North Carolina and 
Dr. Chris Westerfeldt of TUNL for their valuable help in bringing the 
ion implanter back in operation.
We also thank the staff at Yale University's WNSL for 
providing the $^6$Li beam. C.U. thanks the Yale group for their 
nice hospitality during visits to their laboratory.

\newpage 
\bibliography{ne22li6d}

\end{document}